# The effects of oxygen in spinel oxide $Li_{1+x}Ti_{2-x}O_{4-\delta}$ thin films

Yanli Jia[1,2,+], Ge He[1,2,+], Wei Hu[1,2], Hua Yang[1,2], Zhenzhong Yang[1,2], Heshan Yu[1,2], Qinghua Zhang[1,2], Jinan Shi[1,2], Zefeng Lin[1,2], Jie Yuan[1,2], Beiyi Zhu[1,2], Lin Gu[1,2,3], Hong Li[1,2,3] and Kui Jin[1,2,3*]

[1]Beijing National Laboratory for Condensed Matter Physics, Institute of Physics, Chinese Academy of Sciences, Beijing 100190, China.
[2] School of Physical Sciences, University of Chinese Academy of Sciences, Beijing 100049, China
[3]Collaborative Innovation Center of Quantum Matter, Beijing, 100190, China.

**Abstract**

The evolution from superconducting $LiTi_2O_{4-\delta}$ to insulating $Li_4Ti_5O_{12}$ thin films has been studied by precisely adjusting the oxygen pressure during the sample fabrication process. In the superconducting $LiTi_2O_{4-\delta}$ films, with the increase of oxygen pressure, the oxygen vacancies are filled, and the $c$-axis lattice constant decreases gradually. With the increase of the oxygen pressure to a certain critical value, the $c$-axis lattice constant becomes stable, which implies that the $Li_4Ti_5O_{12}$ phase comes into being. The process of oxygen filling is manifested by the angular bright-field images of the scanning transmission electron microscopy techniques. The temperature of magnetoresistance changed from positive and negative shows a non-monotonous behavior with the increase of oxygen pressure. The theoretical explanation of the oxygen effects on the structure and superconductivity of $LiTi_2O_{4-\delta}$ has also been discussed in this work.

**Introduction**

In the research on oxide superconductors, the oxygen always plays an important role in the superconductivity and their normal state behaviors [1, 2]. In copper oxide high-temperature superconductors, such as $Nd_{2-x}Ce_xCuO_{4\pm\delta}$ [3-5], $Pr_{2-x}Ce_xCuO_{4\pm\delta}$ [6-8] and $YBa_2Cu_3O_{7-\delta}$ [9-11], the superconducting transition temperature $T_c$ can be greatly improved in a large range by adjusting the oxygen content during the annealing process, as well as the titanium oxide systems, such as $SrTiO_3$ [12, 13] and TiO [14, 15]. Oxygen has a strong effect not only on superconductivity, but also on many other properties. For instance, the antiferromagnetism [16] and the charge density wave [17] can also be tuned by the oxygen vacancy. Furthermore, the doping and disorder effects induced by oxygen vacancy can cause obviously change on Hall resistance and magnetoresistance (MR) behaviors in the normal state [5, 6, 18]. Studying the oxygen effects is of great help to understand the mechanism, transport and other properties of the oxide superconductor [9, 19, 20].

Among hundreds of spinel oxides, the metallic lithium titanate $LiTi_2O_4$ is the only known oxide superconductor, which $T_c$ is as high as 13.7 K [21], discovered by Johnston *et al*. in 1973 [22]. Previous works have disclosed that $LiTi_2O_4$ is a BCS *s*-wave superconductor with intermediate electron-phonon coupling ($\lambda_{el-ph}$ ~ 0.65) [23, 24]. Nevertheless, an enhanced density of states has been unveiled by magnetic susceptibility [25] and specific heat measurements [23], indicating that *d-d* electronic correlations cannot be ignored in this system. Meanwhile, due to the mixed-valence of Ti ions in the frustrated Ti sublattice, $LiTi_2O_4$ exhibits complicated spin-orbital fluctuations, which is evidenced by the resonant inelastic soft-x-ray scattering [26], nuclear magnetic resonance [27] and magnetic susceptibility measurements [25]. Very recently, electrical transport and tunneling spectra measurements on high quality epitaxial [001]-oriented $LiTi_2O_4$ films have revealed an orbital-related state below ~ 50 K, confirmed by a twofold in-plane angular dependent magnetoresistance, positive MR as well as the relation $\Delta$ ~ -$B^2$ [28].

Interestingly, by tuning the oxygen in the process of sample deposition, the structure of the thin film changes from $LiTi_2O_{4-\delta}$ to $Li_4Ti_5O_{12}$ along with the superconducting-insulation transition [29]. However, this transition seems to happen suddenly, which hinders us from further exploring the nature of the transition. Previous studies on $LiTi_2O_4$ polycrystals have disclosed that the existence of oxygen-site distortion induces prominent changes in the electronic states near $E_F$ [30]. In addition, tunneling experiments on $LiTi_2O_4$ films of different orientations reveal an anisotropic electron-phonon coupling in this system, which is regarded to originate from the *Jahn-Teller* distortions enhanced by oxygen vacancies [31]. Nevertheless, it is still unclear what happens in the microstructure during the transition from $LiTi_2O_4$ to $Li_4Ti_5O_{12}$. Moreover, the mechanism of the oxygen effects on superconductivity of $LiTi_2O_4$ has never been investigated, as well as the transport behaviors in the normal state. Therefore, it is worthy of carrying out a precise tuning on the oxygen content in the process of film deposition to clarify these questions.

In this work, we carefully manipulated the transition from $LiTi_2O_{4-\delta}$ to $Li_4Ti_5O_{12}$ thin films by adjusting the oxygen pressure in the process of pulsed laser deposition (PLD). First, the high quality $LiTi_2O_{4-\delta}$ superconducting thin films can be obtained in the high vacuum environment. Adjusting the oxygen pressure subtly from $10^{-7}$ to $10^{-4}$ Torr, the *c*-axis lattice constant gradually decreases, indicating that the filling of oxygen vacancies dominates in this process. Secondly, the *c*-axis lattice constant stops to decrease with further increasing the oxygen pressure, indicating the finish of transition from $LiTi_2O_{4-\delta}$ to $Li_4Ti_5O_{12}$ phase. These two processes can be revealed from the angular bright-field images (ABF) of LiTi$_2$O$_{4-\delta}$ and Li$_4$Ti$_5$O$_{12}$ by the scanning transmission electron microscopy (STEM) techniques. In addition, the temperature ($T_{ch}$) of MR from positive to negative shows a nonmonotonic behavior, *i.e.* first decrease then increase, with the increase of oxygen pressure. Combined with the electron energy-loss spectroscopy (EELS) measurements, we suggest that the decrease of $T_{ch}$ under lower oxygen pressure stems from the suppression of orbit-related state via filling the oxygen vacancies, and the increase of $T_{ch}$ under higher oxygen pressure is due to the appearance of grain boundaries between LiTi$_2$O$_{4-\delta}$ and Li$_4$Ti$_5$O$_{12}$ phases, which dominates the positive MR (p-MR).

**Experiments**

The (00*l*)-oriented Li$_{1+x}$Ti$_{2-x}$O$_{4-\delta}$ (0 ≤ x ≤ 1/3) thin films were grown on (00*l*) MgAl$_2$O$_4$ (MAO) substrates by PLD with a $K_r$F excimer laser (λ = 248 nm). Before the deposition, the MAO substrates were annealed at 1000 °C for 5 hours in the air [32, 33] to obtain the smooth surface. The sintered Li$_4$Ti$_5$O$_{12}$ ceramic target was used to fabricate the films, with pulse frequency of 4 Hz, energy density of 1.5 J/cm$^2$, and the deposition temperature of ~ 700 °C. The deposition rate was determined by measuring the thickness of ultra-thin films using X-Ray Reflectivity (XRR) analysis. In this study, we fixed the film thickness ~ 150 nm. After the deposition, all the thin films were quenched to the room temperature in situ.

X-ray diffraction (XRD) was employed to characterize the phase and crystalline quality of Li$_{1+x}$Ti$_{2-x}$O$_{4-\delta}$ (0 ≤ x ≤ 1/3) thin films. The microstructure was detected by the spherical aberration-corrected scanning transmission electron microscopy techniques (Cs-STEM). The transport properties were measured by the Quantum Design Physical Property Measurement System (PPMS) with the temperature down to 2 K and magnetic field up to 9 T. Samples were etched into Hall bar by the UV lithography and Ar plasma etching technology to measure the resistance properties.

**Results and Discussion**

The *θ*-2*θ* XRD spectra of (001) Li$_{1+x}$Ti$_{2-x}$O$_{4-\delta}$ (0 ≤ x ≤ 1/3) samples grown with different oxygen pressures ($P_{O2}$) are shown in Fig. 1(a). The (001)-oriented LiTi$_2$O$_{4-\delta}$ thin films are achieved when the films are deposited under $P_{O2}$ ≤ 10$^{-6}$ Torr. Instead, the (001)-oriented Li$_4$Ti$_5$O$_{12}$ thin films are formed at $P_{O2}$ > 10$^{-4}$ Torr. The XRD patterns of the samples in different oxygen pressures look similar except that the diffraction peaks gradually shift to higher angle in the LiTi$_2$O$_{4-\delta}$ films at larger $P_{O2}$. In order to check the quality of the thin films, we also performed *φ*-scan. In Fig. 1(b), the *φ*-scans of (404) plane of both LiTi$_2$O$_{4-\delta}$ and Li$_4$Ti$_5$O$_{12}$ samples display four-fold symmetry with uniformly distributed peaks. From the $\theta$-$2\theta$ XRD spectra, we can extract the value of the out-of-plane lattice constant (*c-axis*) as a function of the oxygen pressure. As seen in Fig. 1(c), when $P_{O2} < 10^{-4}$ Torr, $c$ gradually decreases with increasing $P_{O2}$. However, when $P_{O2}$ is higher than $10^{-4}$ Torr, $c$ is saturated, indicating the formation of $Li_4Ti_5O_{12}$ phase. As a result, a phase transition from $LiTi_2O_{4-\delta}$ to $Li_4Ti_5O_{12}$ has been successfully achieved solely by adjusting $P_{O2}$ during the sample deposition.

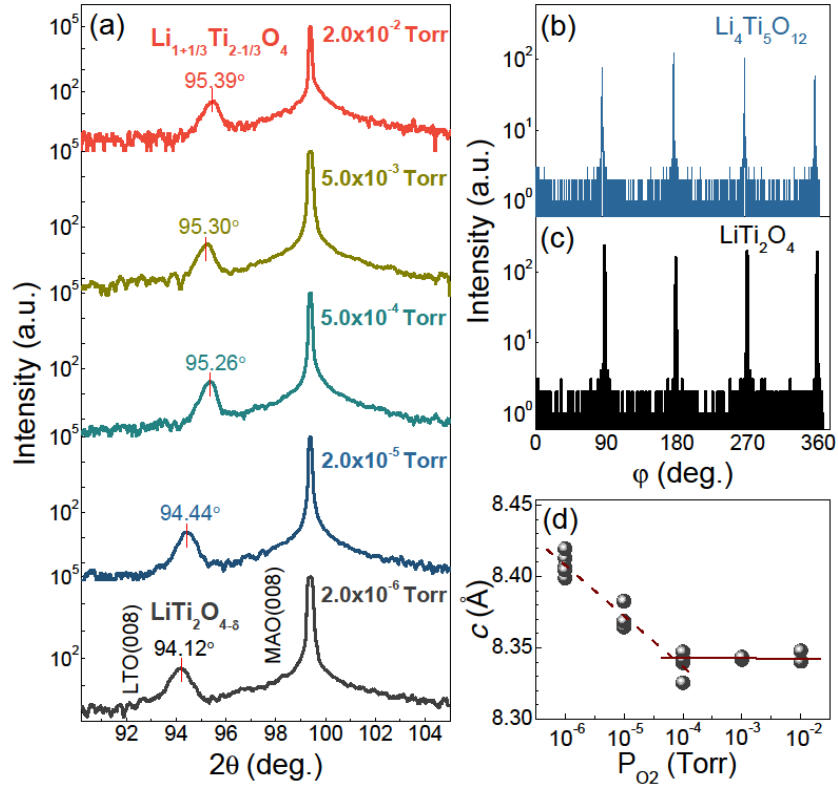

FIG.1 (Color online) (a) The $\theta$-$2\theta$ XRD spectra of epitaxial $Li_{1+x}Ti_{2-x}O_{4-\delta}$ (0 ≤ x ≤ 1/3) thin films grown on (001) MAO substrates at different $P_{O2}$. (b) The $\varphi$-scan measurements of $LiTi_2O_{4-\delta}$ and $Li_4Ti_5O_{12}$ thin films on MAO (001) in the (404) reflection. (c) The lattice constant along $c$ axis of the samples versus $P_{O2}$.

In order to further study the effects of oxygen on superconducting state and normal state, we systematically measured the resistance ($R$) properties of various thin films from $LiTi_2O_{4-\delta}$ to $Li_4Ti_5O_{12}$. The $R$-$T$ curves of the $LiTi_2O_{4-\delta}$ thin films with different oxygen pressures are shown in Fig. 2(a). Increasing the oxygen pressure during the deposition, the samples experience a transition from metal to semiconductor in the normal state. The $Li_4Ti_5O_{12}$ is insulating, whereas the $LiTi_2O_{4-\delta}$ is metallic and become the superconductor at a lower temperature. The $T_c$ of the $LiTi_2O_{4-\delta}$ thin films is quite robust against the $P_{O2}$. In Fig. 2(b), the residual resistivity ratio (*RRR*) decreases monotonically with increasing $P_{O2}$. Here, the RRR is defined as the ratio of room temperature resistivity to the resistivity right before entering the superconducting state, *i.e.* $R(300K)/R(T_c^{onset})$.

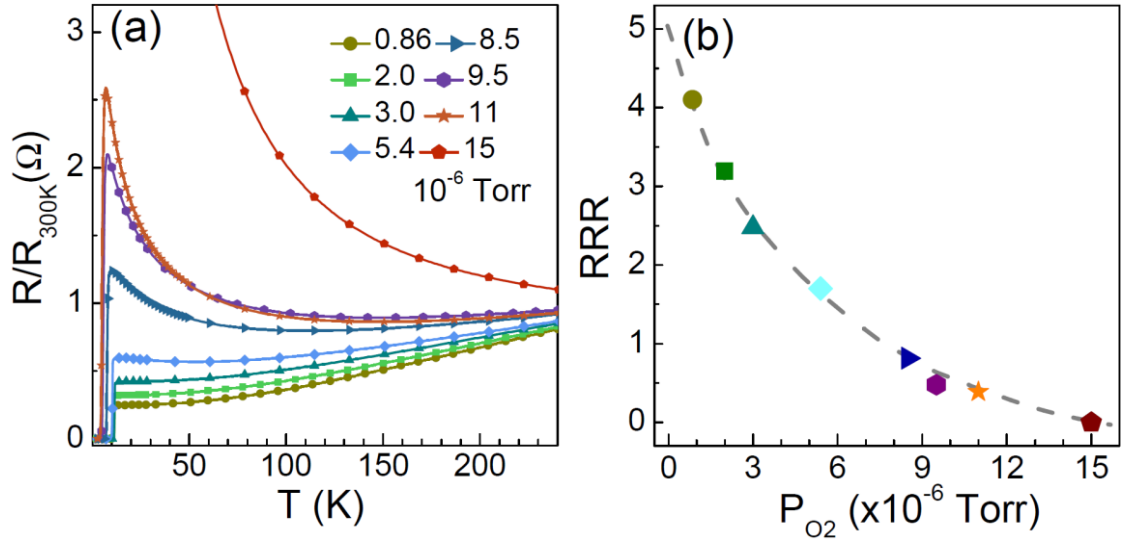

FIG.2 (Color online) (a) The *R*-T curves of $Li_{1+x}Ti_{2-x}O_{4-\delta}$ ($0 \leq x \leq 1/3$) thin films grown on (001) MAO substrate with different oxygen pressures in depositions. (b) The *RRR* decreases monotonically with the increase of oxygen pressure, which is defined as R (300K)/R ($T_c^{onset}$).

To figure out the evolution of the microstructure from $LiTi_2O_{4-\delta}$ to $Li_4Ti_5O_{12}$, we have carried out STEM measurements on these high-quality samples. Figure 3 shows the $LiTi_2O_{4-\delta}$ lattice and the corresponding ABF STEM images along the [110] direction. Typical ABF images on different regions of the $LiTi_2O_{4-\delta}$ sample disclose the ideal structure (Fig. 3(a)) and regions with oxygen vacancies (Figures 3b and 3c), which are imaged as light gray spots. The positions of oxygen atoms can be divided into two types, *i.e.* $O_1$ and $O_2$ (see Fig. 3(d)). The stacking density of oxygen at $O_1$ is higher than that at $O_2$ as seen from the [110] direction, which can be seen in the lower panel of Fig. 3(e). In different regions, the contrast between $O_1$ and $O_2$ is varying as seen in Fig. 3(e) and 3(f), clarifying apparent oxygen vacancies. However, such oxygen vacancies have not been observed in the $Li_4Ti_5O_{12}$ samples [34]. It is known that the $LiTi_2O_{4-\delta}$ exhibits serious aging effects in forms of polycrystal and single crystal [30]. The $LiTi_2O_{4-\delta}$ thin films, especially the one deposited in the higher vacuum, are much more stable. It is reasonable to speculate that the samples in higher vacuum will contain more oxygen vacancies. Increasing oxygen pressure will cure these oxygen vacancies and finally turn the superconducting phase to insulating $Li_4Ti_5O_{12}$.

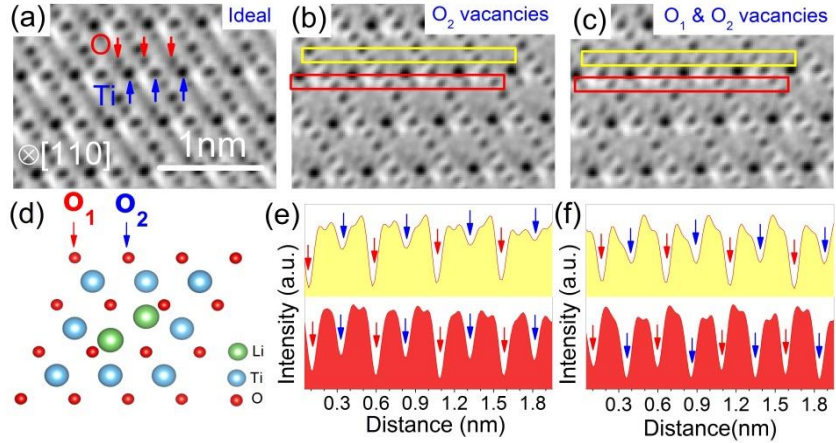

FIG. 3 (Color online) (a)-(c) The ABF image of $LiTi_2O_{4-\delta}$ thin film with (a) ideal structure, (b) $O_2$ vacancy area and (c) $O_1$ & $O_2$ vacancy areas, where the yellow and the red rectangle areas represent the regions of the oxygen atoms which are arranged in the upper and lower layers of Ti sites, respectively. (d) The illustration of lattice structures of $LiTi_2O_4$ along [110] direction. (e)-(f) The energy profiles corresponding to the central horizontal lines in the rectangle areas in (b) and (c). Note that the lower panel in (e) exhibits contrast between $O_1$ (red arrows) and $O_2$ (blue arrows) close to the ideal structure.

The phase evolution with $P_{O2}$ should inevitably make difference in the electronic states. In $LiTi_2O_{4-\delta}$, one concern is about the orbital-related state, which has been unveiled in previous work but remains unclear in the origin [25]. One of the evidence is the crossover from negative MR to positive MR at $T_{ch}$ ~ 50 K in the normal state. Entering the superconducting state, the orbital-related state interacts with Cooper pairs and results in an unexpected relation between the superconducting gap and the applied magnetic field, i.e. $\Delta \sim -B^2$. Such relation implies the coexistence of the superconducting state and the orbital-related state. Therefore, the next question is how the oxygen makes the influence on these two states.

To clarify this issue, we finely tune the oxygen pressure around $5.0 \times 10^{-6}$ Torr to avoid the $Li_4Ti_5O_{12}$ phase. Then, we focus on the effects of $P_{O2}$ on resistance and magnetoresistance. In such fine tuning process, the vacuum reading is not a good coordinate due to the limitation of the vacuum gauge. Fortunately, the RRR decreases monotonically with the increase of oxygen pressure, which can reflect the trend of $P_{O2}$ and the oxygen defects as discussed above. Thus, we use RRR to index the samples, namely S1 to S8 with RRR in the range between 5.6 and 1.5. As shown in Fig. 4(a), the $T_c$ seems unchanged in the tuning range. For samples S1 to S5, the MR at 35K changes from positive to negative as seen in Fig. 4(b). By fitting the MR with the Kohler's formula, i.e. MR ~ $A_0B^2$, the slope $A_0$ can be obtained for these samples at different temperatures. In Drude model, $A_0$ is proportional to $\mu^2$ (i.e. $\mu = e\tau/m$) with $\mu$ the mobility, $\tau$ the relaxation time and $m$ the electron mass. With increasing the temperature, the value of $A_0$ decreases from positive to negative (see Fig. 4(c)). A negative mobility cannot be understood in this simplified model, and the negative MR is interpreted as the suppression of spin-orbital fluctuations in this system [28]. The crossing temperature ($T_{ch}$) from p-MR to n-MR is extracted from Fig. 4(c) and plotted in Fig. 4(d). For S1 to S5, the higher is the $T_{ch}$, the bigger is the $RRR$.

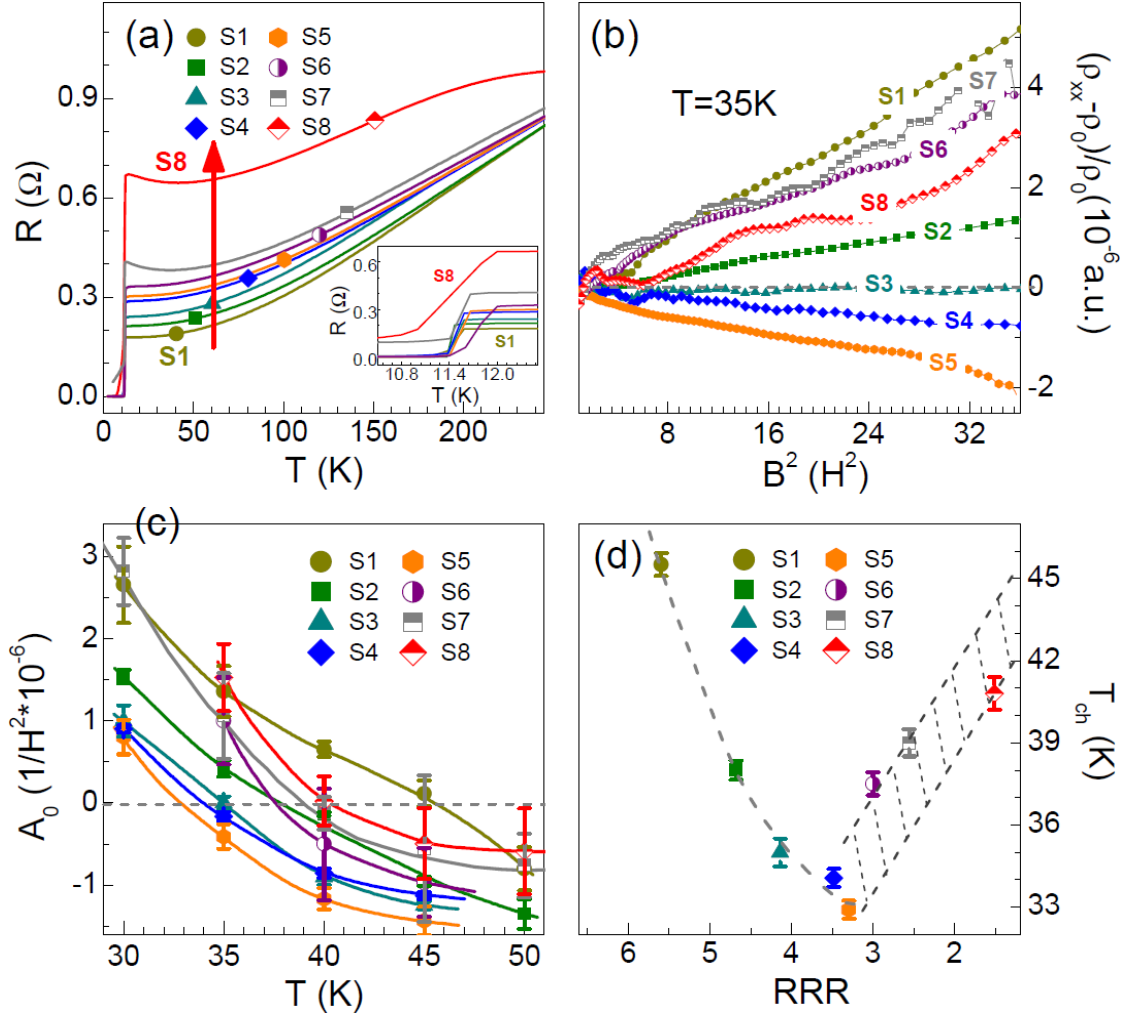

FIG.4 (Color online) (a) The $R$-T curves with increase of oxygen pressures are defined as S1-S8 in sequence. The inset is the magnification of $R$-T curves. (b) The field dependence of the in-plane MR of S1-S8 grown on (001) MAO substrates at 35K. (c) The slope value $A_0$ of in-plane MR can be obtained for various samples at different temperatures. With temperature increasing, the value of $A_0$ changes from positive to negative. (d) The relationship between the $RRR$ and $T_{ch}$, and $T_{ch}$ show a nonmonotonic behavior with the decrease of $RRR$.

However, further reducing the $RRR$, this relation does not hold. For instance, the positive MR becomes stronger for the samples deposited under higher oxygen pressure, *e.g.* S6-S8, and thus the $T_{ch}$ goes up. In this regime, the formation of $Li_4Ti_5O_{12}$ phase may lead to more boundaries in phase separated samples. Such inhomogeneity in the magnetic field usually exhibits strong p-MR [35, 36]. For the samples S1-S5, the p-MR below $T_{ch}$ mainly origins from the orbital-related state since the $LiTi_2O_{4-\delta}$ phase dominates the transport [25]. We speculate that filling the oxygen vacancies seems to suppress the positive MR and thus the orbital-related state.

In order to verify this assumption, we should evaluate the effects of oxygen vacancies on Ti valance. Although oxygen vacancies have been detected by STEM, the content of oxygen vacancies cannot be quantified. Therefore, we collected EELS profiles of both LiTi$_2$O$_{4-\delta}$ and Li$_4$Ti$_5$O$_{12}$ films. As seen in Fig. 5, the Ti $L_{2,3}$ edges, from $2p_{1/2}$ and $2p_{3/2}$ to $3d$ orbits respectively, split into two peaks in Li$_4$Ti$_5$O$_{12}$, but not in LiTi$_2$O$_{4-\delta}$. Usually, the splitting of $L_{2,3}$ is attributed to the degeneracy lifting of Ti $3d$ orbits by the crystal field.

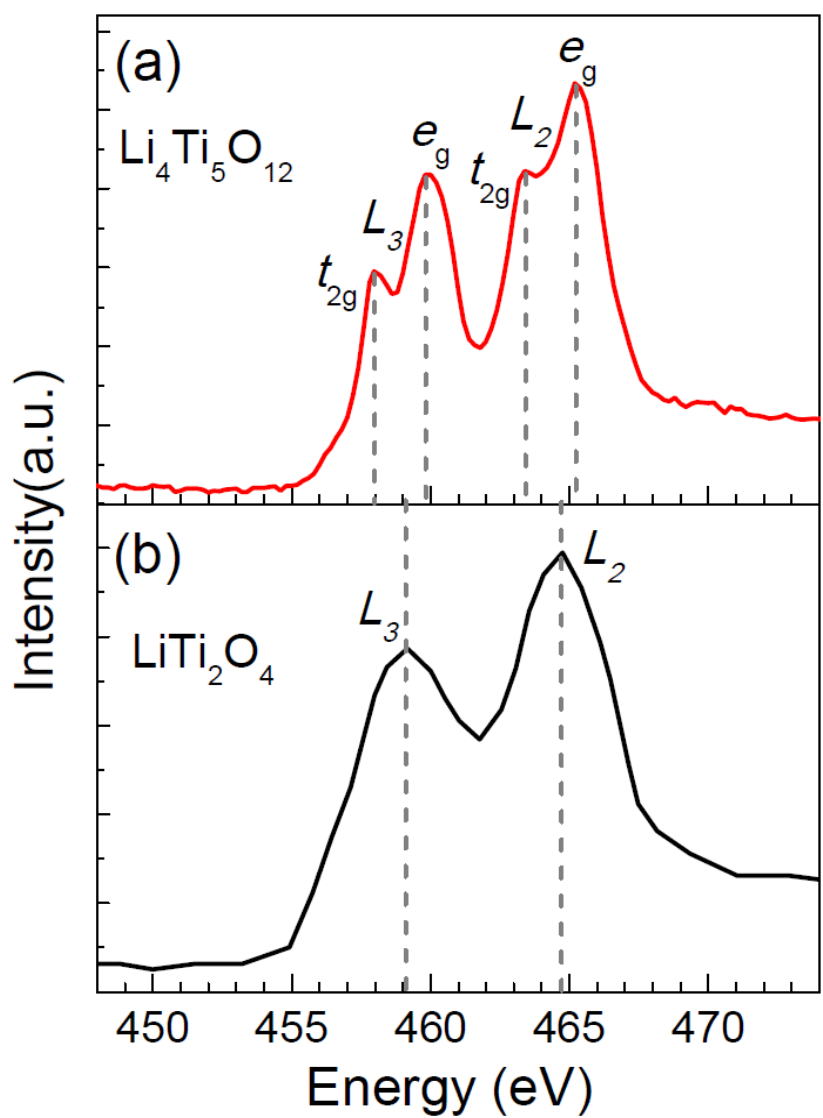

FIG. 5 (Color online) (a) The EELS profiles for Ti $L_{2,3}$ edges of the Li$_4$Ti$_5$O$_{12}$ thin film. The $L_2$ and $L_3$ edges split into four peaks. (b) The EELS profiles for Ti $L_{2,3}$ edges of the LiTi$_2$O$_4$ thin film. Only two peaks appear of $L_{2,3}$ edges.

The missing split of peaks in LiTi$_2$O$_{4-\delta}$ EELS may origin from two reasons. First, the energy gaps between $e_g$ and $t_{2g}$, named $\Delta_{e-t}$, is too small to be discernable in EELS. According to the band calculations, $\Delta_{e-t}$ equals to 2.1 eV and 2.4 eV for the ideal structure LiTi$_2$O$_4$ and Li$_4$Ti$_5$O$_{12}$, respectively [37]. Moreover, four peaks were also observed in Li$_7$Ti$_5$O$_{12}$, where $\Delta_{e-t}$ equals to 1.8 eV [34]. Considering the existence of oxygen vacancies, which may further distort the Ti-O octahedrons, we do not expect a smaller crystal field. Therefore, the change of $\Delta_{e-t}$ cannot account for the discernable peak splitting in LiTi$_2$O$_{4-\delta}$. Secondly, the valance of Ti in ideal LiTi$_2$O$_4$ is +3.5. If large numbers of oxygen vacancies exist in LiTi$_2$O$_{4-\delta}$, the Ti$^{3.5+}$ will move towards Ti$^{3+}$. In this condition, the electrons on $t_{2g}$ band increase, and thereby the hoping possibility from Ti $2p$ to $t_{2g}$ is reduced due to the Pauli exclusion principle. Consequently, oxygen vacancies will smear out the peaks of Ti $2p$ to $t_{2g}$ in EELS.

Based on the EELS results, we can give a reasonable explanation for the suppression of the orbital-related state by filling oxygen vacancies. In general, the formation of the orbital order results from the band split near the Fermi level. As for LiTi$_2$O$_4$, crystal field splits Ti $3d$ bands to e$_g$ and t$_{2g}$ bands [38]. The oxygen vacancies in LiTi$_2$O$_{4-\delta}$ system, on the one hand, enhance the distortion of Ti-O octahedrons, on the other hand, dope electrons to enhance the electron correlations, which are beneficial to the formation of the orbital-related state. With the filling of oxygen vacancies, the valence of Ti decreases and some of Ti sites become empty states, which will weaken the orbital-related state.

Compared to the obviously suppressed orbital-related state, the $T_c$ of the samples is almost unchanged. Actually, the O $2p$ bands are far below the Fermi level with weak $p$-$d$ hybridizations [30, 38]. Although the oxygen vacancies lift up the Fermi level by the doping effect and influence on the splitting of Ti $3d$ bands by the crystal field, the density of states near Fermi surface does not change, and thus the $T_c$ holds.

In conclusion, we studied the evolution from LiTi$_2$O$_{4-\delta}$ to Li$_4$Ti$_5$O$_{12}$ with increasing oxygen pressure during the thin film deposition. By transport and STEM measurements, we have disclosed that there are two processes happening during the evolution, *i.e.* the filling of oxygen vacancies which weakens the orbital-related state and the forming of Li$_4$Ti$_5$O$_{12}$. The EELS results of the LiTi$_2$O$_{4-\delta}$ and Li$_4$Ti$_5$O$_{12}$ samples provide the evidence that the orbital-related state is suppressed by the filling of oxygen vacancies. The evolution of electronic states by adjusting the oxygen content thus gives an insight into the interaction between the orbital-related state and the superconductivity in LiTi$_2$O$_{4-\delta}$.

## Acknowledgments

We thank K. Liu for fruitful discussions and L. H. Yang for technique support. This work was supported by the National Key Basic Research Program of China (2015CB921000, 2016YFA0300301, 2017YFA0303003), the National Natural Science Foundation of China (11674374, 11474338, 11574372), the Beijing Municipal Science and Technology Project (Z161100002116011, D161100002416003), the Key Research Program of Frontier Sciences, CAS (QYZDB-SSW-SLH001, QYZDB-SSW-SLH008), the Open Research Foundation of Wuhan National High Magnetic Field Center (PHMFF2015008), and the Strategic Priority Research Program of the CAS (XDB07020100).